\documentclass{PoS}

\title{A Generic Algorithm for IACT Optical Efficiency Calibration using Muons}

\ShortTitle{A Generic Algorithm for IACT Muon Calibration}

\author{A.M.W. Mitchell\speaker{}\thanks{Member of the International Max Planck Research School for Astronomy and Cosmic Physics at the University of Heidelberg (IMPRS-HD) and the Heidelberg Graduate School of Fundamental Physics (HGSFP).}, V. Marandon, R. D. Parsons, for the H.E.S.S. Collaboration\\
       Max-Planck-Institut f\"{u}r Kernphysik, P.O. Box 103980, D 69029 Heidelberg, Germany\\
        E-mail: \email{Alison.Mitchell@mpi-hd.mpg.de}}

\abstract{Muons produced in Extensive Air Showers (EAS) generate ring-like images in Imaging Atmospheric Cherenkov Telescopes when travelling near parallel to the optical axis. From geometrical parameters of these images, the absolute amount of light emitted may be calculated analytically. Comparing the amount of light recorded in these images to expectation is a well established technique for telescope optical efficiency calibration. However, this calculation is usually performed under the assumption of an approximately circular telescope mirror. The H.E.S.S. experiment entered its second phase in 2012, with the addition of a fifth telescope with a non-circular 600\,m$^2$ mirror. Due to the differing mirror shape of this telescope to the original four H.E.S.S. telescopes, adaptations to the standard muon calibration were required. We present a generalised muon calibration procedure, adaptable to telescopes of differing shapes and sizes, and demonstrate its performance on the H.E.S.S. II array.}

\FullConference{The 34th International Cosmic Ray Conference,\\
		30 July- 6 August, 2015\\
		The Hague, The Netherlands}

\begin{document}
\section{Introduction} 
\label{sec:intro}
\noindent Charged particles within Extensive Air Showers (EAS), generated by energetic $\gamma$-rays or Cosmic Rays penetrating the Earth's atmosphere, travel faster than the local speed of light, thereby producing Cherenkov radiation, subsequently detectable by Imaging Atmospheric Cherenkov Telescopes (IACTs). In order to record snapshots of EAS, fast imaging cameras are employed, with typical integration times of 16\,ns. The nature and energy of the primary particle may be determined from the shape and intensity in photoelectrons of the recorded images respectively.  The combined optical efficiency of the various telescope and camera components directly influences the image intensity and can significantly affect the energy reconstruction. 

\noindent Muons produced in hadronic EAS are more penetrating than lighter leptons, and may scatter to travel further from the shower axis; this reduces  the amount of radiation from parent hadronic EAS recorded in the same image as a muon. The ring-shaped images generated in the camera are easily identifiable; the use of these images for optical efficiency calibration is a well-established method within the IACT community \cite{bib:Vacanti94}. \noindent Comparing the recorded image amplitude for a given muon ring to the theoretical prediction provides a measure of the light collection efficiency of the telescope system. The mean of this parameter distribution over a given time period is termed the muon efficiency, $\varepsilon_\mu$. 

\noindent Within the analysis chain, this muon efficiency is used to correct energy estimates by a factor accounting for the discrepancy between the current state of the system and the relevant Monte Carlo (MC) reference state. The optical efficiency can also have a large effect on the Instrument Response Functions which must be accounted for by regular calibration. 
 
\noindent H.E.S.S. (High Energy Stereoscopic System) is an array of IACTs situated at 1800m altitude in the Khomas Highlands in Namibia \cite{bib:Hinton04}, enhanced in 2012 by the addition of a fifth telescope of 600m$^2$ mirror area (hereafter CT5). This enables the energy range of H.E.S.S. to be extended at the low end, pushing down the energy threshold to $\sim O(10\,\mathrm{GeV})$. Whilst the muon calibration technique has been utilised for telescopes of the H.E.S.S. array since its inception in 2003 \cite{bib:Bolz04,bib:Leroy04}, standard assumptions of the muon reconstruction procedure, in particular a circular mirror shape, were no longer valid for CT5. Here, the adaptations subsequently made to the standard muon calibration are outlined, which render the algorithm suitable to varying IACT mirror configurations. Improved performance over previous implementations of this calibration procedure is also demonstrated. 

\section{Muon Calibration Procedure}
\subsection{Muon Image Identification}
\label{sec:muonimageid}

\noindent The raw image is initially cleaned using tail-cuts (dual-threshold: all pixels above an upper threshold and neighbouring pixels above a lower threshold are kept), with upper and lower thresholds of 7 and 5 photoelectrons respectively. An image is subsequently classified as ring-like if a circle is found using the Chaudhuri-Kundu formula \cite{bib:Chaudhuri93}. In this case, the circle fit is re-performed using the results of the first fit as seed parameters and excluding pixels lying further than 1.5 pixel widths away from the seed ring radius. This helps to refine the ring parameters (radius, centre position, ring width) by reducing fit biases due to scattered light from parent hadronic showers. 
Intensity information for ``on ring'' pixels  (pixels within 0.5$^{\circ}$ radial distance either side of the fitted ring radius) is then taken from an uncleaned version of the image. 

\noindent A set of cuts optimised for this algorithm and for different telescope types, are used to select muon images for calibration. These cuts ensure that the selected ring images are (i) well above threshold, (ii) well contained, (iii) ring-like and (iv) that the influence of hardware issues is minimal. 
Pixel related cuts are made on the total number, N$_{\mathrm{pix}}$ (reason i),  the number in the outermost layer at the camera edge, N$_{\mathrm{EdgePix}}$ (reason ii), the average number of neighbouring pixels, $\langle\mathrm{NN}\rangle$ (reason iii), and the number classified as broken, N$_{bp}$ (reason iv). Additional cuts are made on the ring radius, $r$ (reason i) and on the outermost radius of the ring, $r + \rho_r$ (reason ii), where $\rho_r$ is the ring impact parameter (the distance from the centre of the ring to the centre of the camera).

\subsection{Predicted Image Intensity}
\label{sec:prediction}

\noindent The total intensity $I_{pe}$ of Cherenkov light produced by a muon as a function of azimuthal angle $\phi$ and impact parameter $\rho$, the distance from the muon impact position on the mirror to the centre of the mirror, is given by:

\begin{equation}
I_{pe} = \varepsilon_\mu I(\theta_c,\rho,\phi,\omega) = \varepsilon_\mu \frac{\alpha}{2} \frac{\omega}{\theta_c}I\sin(2\theta_c)D(\rho,\phi)
\label{eq:Intensitycalc}
\end{equation}
where $\omega$ is the angular camera pixel size; $\alpha$ the fine structure constant; and $\varepsilon_\mu$ the muon efficiency, a global scaling parameter relating the total light intensity to the expectation, determined by a log-likelihood based minimisation procedure. The Cherenkov opening angle of light emission, $\theta_c$, corresponds to the ring radius, $r$ as measured in degrees in the camera frame; and $D(\rho,\phi)$ is the distance across the mirror along which the collected light must be integrated. 
The intrinsic intensity $I = \int \lambda^{-2} \mathrm{d}\lambda$ is integrated over optical wavelengths $300\,\mathrm{nm} < \lambda < 600\,\mathrm{nm}$, where the atmospheric refractive index is assumed to be $\lambda$ independent \cite{bib:Bolz04}. 

 \begin{figure*}[t]
  \centering
  \includegraphics[width=\textwidth]{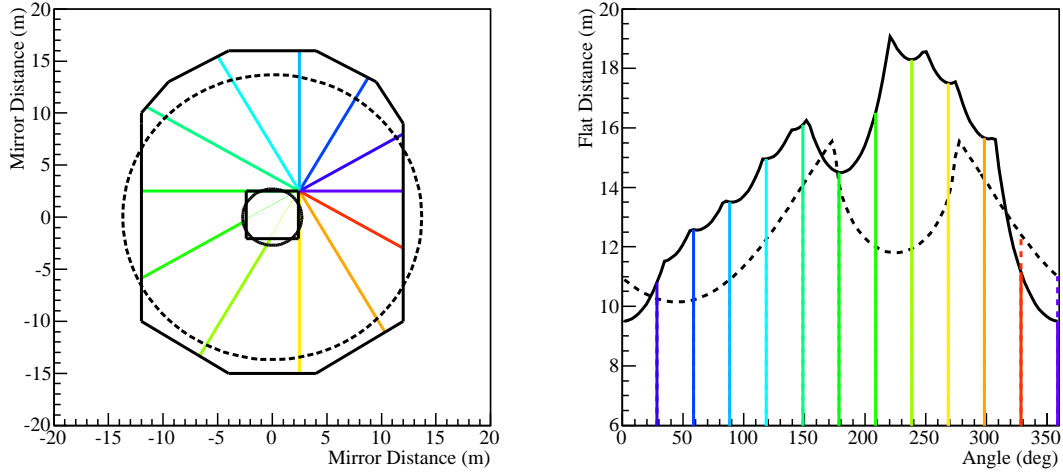}
  \caption{Line integration along the mirror surface as a function of azimuth angle for a given impact position, shown for CT5 with the true mirror profile (solid line) and under a circular approximation (dashed line).}
  \label{fig:CT5mircolourprofile}
 \end{figure*}

\noindent Previously, $D(\rho,\phi)$ was determined analytically assuming a circular mirror dish for H.E.S.S. I; an approximation no longer adequate for describing the CT5 mirror (figure \ref{fig:CT5mircolourprofile}). Instead, $D(\rho,\phi)$ is obtained by interpolating the distance between a set of points describing the outside of the dish. 
Figure \ref{fig:CT5mircolourprofile} shows that the variation in distance with azimuth angle is more accurately described by the interpolation, avoiding biases due to a circular approximation, such as an apparent shift in azimuthal position of the central hole. The azimuthal muon ring intensity profile (\ref{eq:Intensitycalc}) takes the same shape as that shown in figure \ref{fig:CT5mircolourprofile} for $D(\rho,\phi)$ since all other parameters are $\phi$ independent.

\begin{figure*}[b]
	\begin{minipage}{0.48\textwidth}
	\centering
	\includegraphics[width = \textwidth]{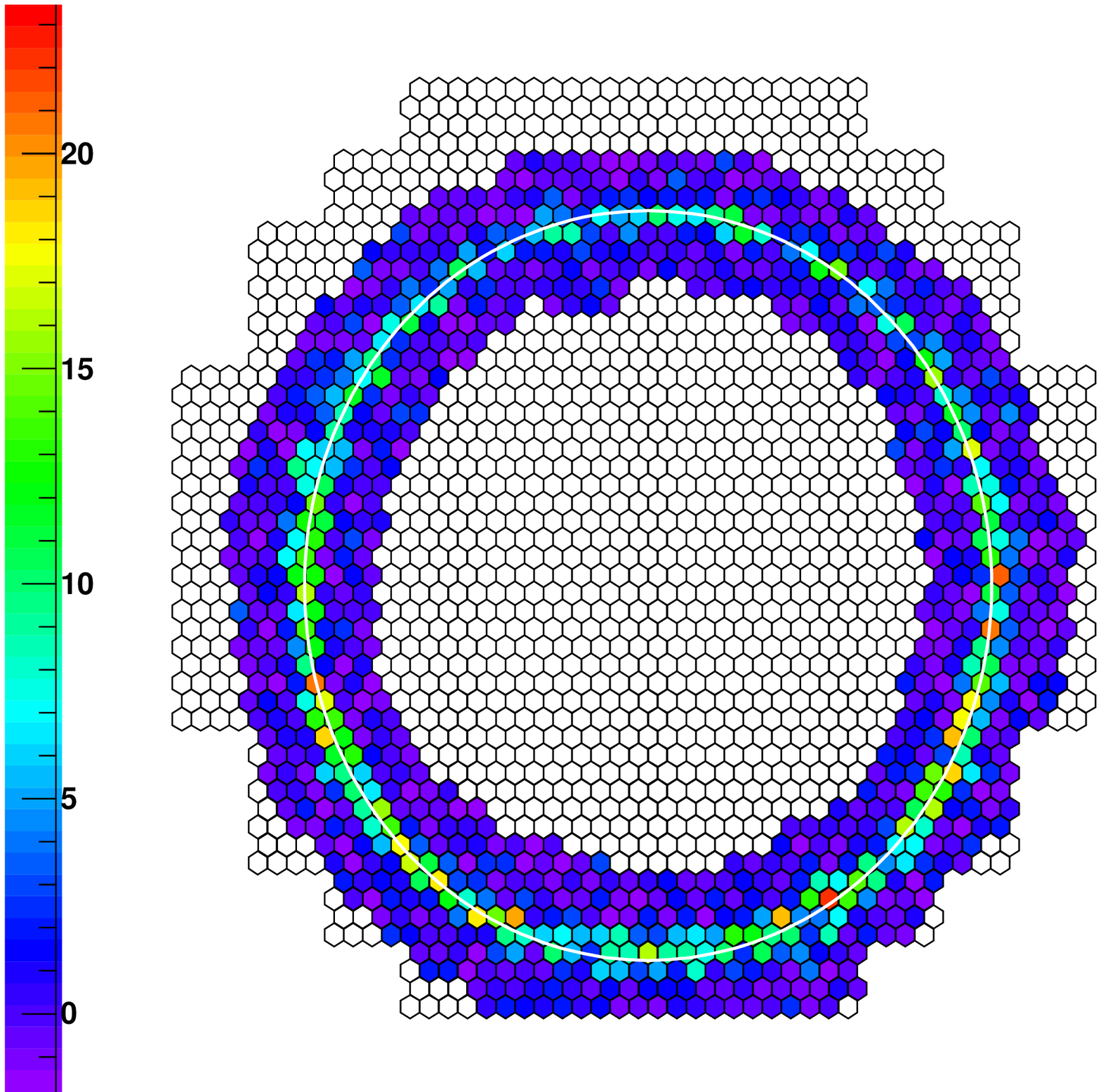}
	\end{minipage}
	\hspace{12pt}
	\begin{minipage}{0.48\textwidth}
	\centering
	\includegraphics[width = \textwidth]{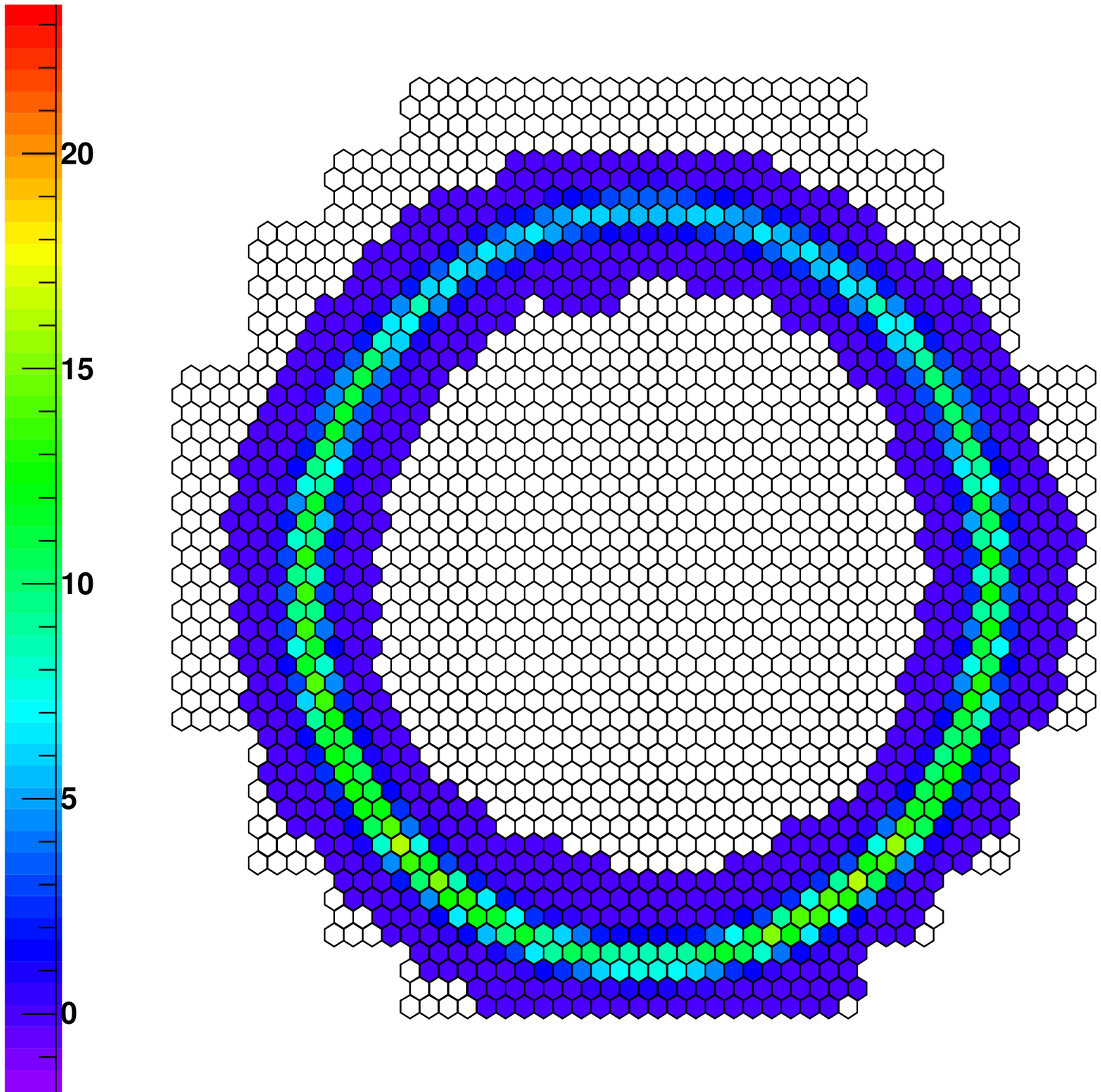}\\
	\end{minipage}
    \caption{Muon event as seen in the CT5 camera (left) and the expectation from the 2D fit (right). Colour scale corresponds to the charge in photoelectrons per pixel after calibration. Statistical fluctuations may yield negative values after baseline subtraction. }
    \label{fig:MuonCamera}
\end{figure*}

\subsection{Muon Efficiency Determination with a 2D Fit}
\label{sec:2deff}
\noindent The raw pixel-wise intensities for all ``on ring'' pixels are summed along each azimuthal direction around the ring, producing a 1D distribution of intensity as a function of azimuth angle, $I(\phi)$.  This raw intensity distribution is smoothed around the ring with a four pixel-widths moving average sliding window. After smoothing, the 1D intensity distribution is broadened radially, according to a Gaussian distribution centered on the ring location, and spread by the fitted ring width. In all stages, the total integral charge is preserved. 
 The 2D smoothed image is fitted with a 2D model using a pixelwise log-likelihood, with the overall efficiency, impact position on the mirror and ring width as free parameters of the fit (the ring centre and ring radius are kept as fixed variables). 
In contrast, the algorithm previously implemented used a 1D $\chi^2$ fit to the azimuthal ring intensity profile to determine $\varepsilon_\mu$.
The pixel likelihood is calculated from a Gaussian approximation after \cite{bib:deNaurois09}, for the case of high model prediction compared to the observed signal. An example muon event recorded by CT5 and the 2D model expectation is shown in figure \ref{fig:MuonCamera}.

\noindent Events passing further cuts on the impact parameter (distance from the mirror centre) and on the ring width, (for reasons ii and iii introduced in section \ref{sec:muonimageid}) are retained for the calibration. The telescope-wise muon efficiency is given by the mean of the distribution from single events per run (approx. half an hour of observations). The obtained muon efficiency is compared to a reference found by performing the muon analysis on MC. This reference MC uses input telescope optical efficiencies  corresponding to the set of lookup tables used for the energy reconstruction. The ratio of the reconstructed MC muon efficiency to the muon efficiency obtained on data defines an optical efficiency correction factor, applied to energy estimates from individual telescopes at analysis level.

\section{Performance}
\label{sec:performance}
\begin{figure*}
	\begin{center}
	\vspace{-5mm}
	\begin{minipage}{0.48\textwidth}
	\centering
  	\includegraphics[width=\textwidth]{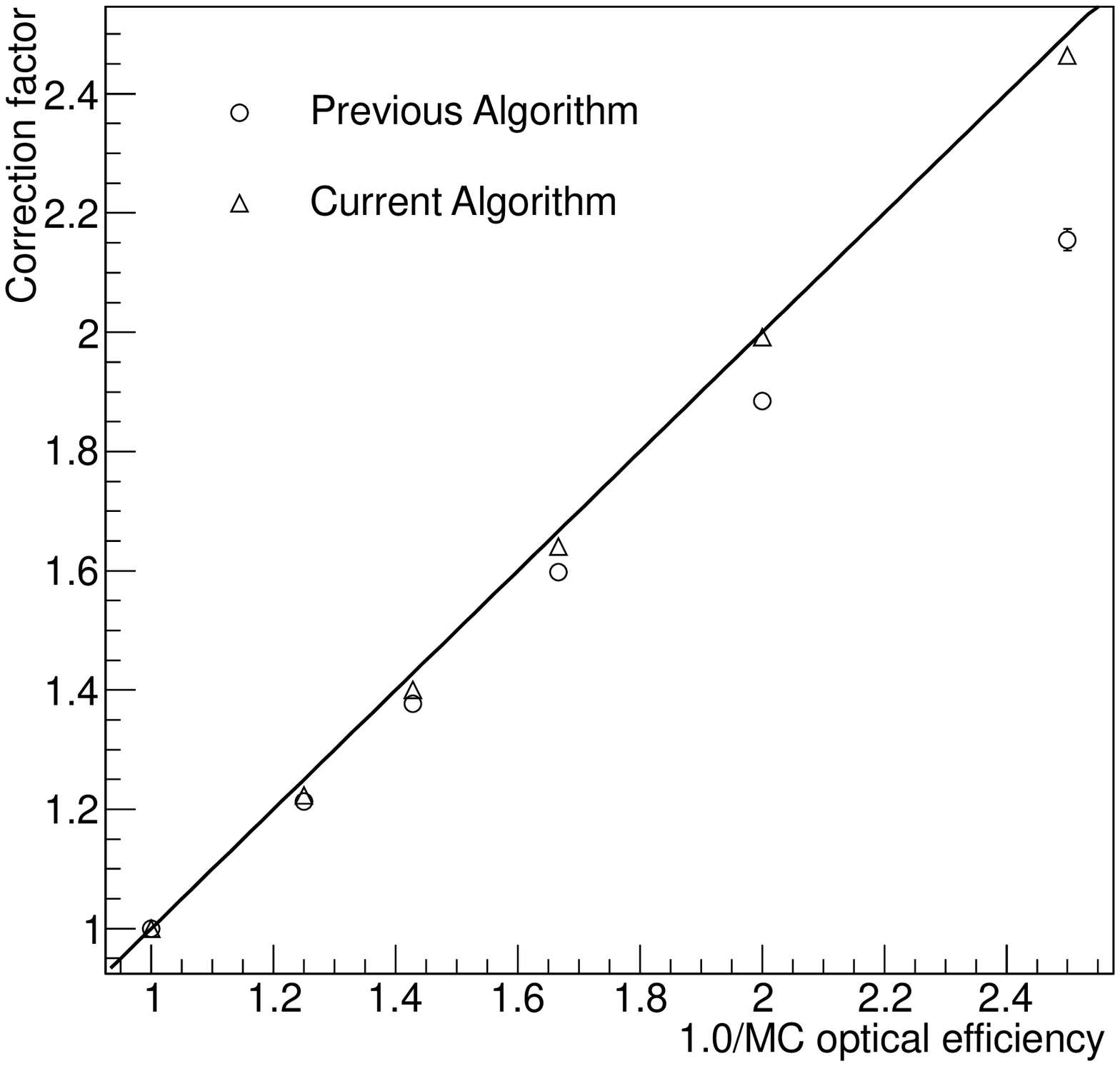}
	 \label{fig:phase1blinearity}	
	\end{minipage}
  \hspace{10pt}
 \begin{minipage}{0.48\textwidth}
  \centering
  \includegraphics[width=\textwidth]{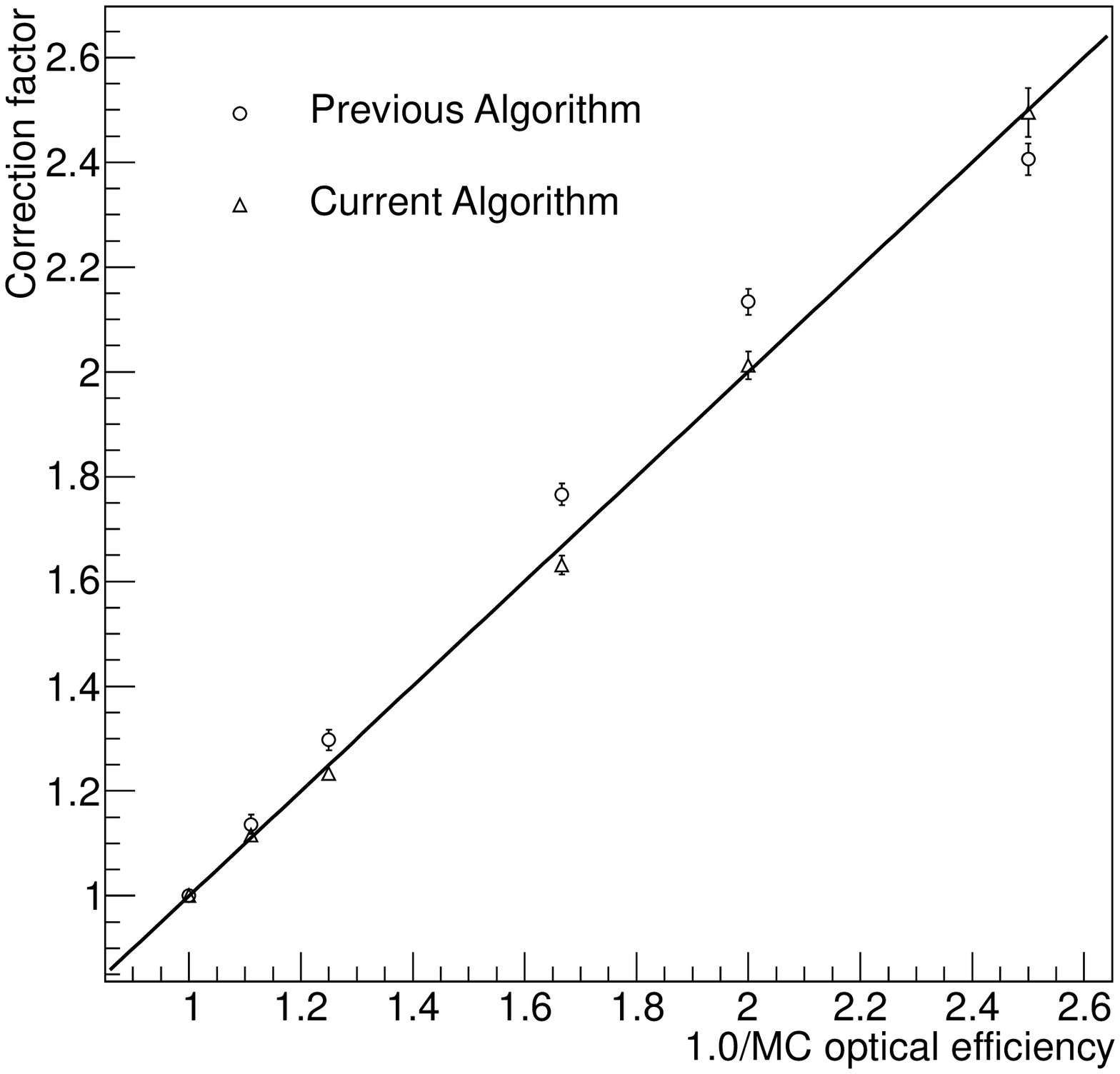}
  \label{fig:phase2a80linearity} 
 \end{minipage}
 \vspace{-5mm}
  \caption{Linearity of previous and new algorithms for both the H.E.S.S. I telescopes (left) and for CT5 (right). In both cases, an improvement over the performance of the previous algorithm is seen.}  	
	\end{center}
  	\label{fig:mclinearity}
\end{figure*}

\noindent Two major changes to the muon calibration algorithm in the H.E.S.S. analysis chain, the arbitrary mirror description and the 2D pixel log-likelihood fitting, have led to demonstrable improvements in the performance. Ideally, the correction factor should be minimally biased by efficiency degradation. This can be checked by performing the muon calibration on MC of various (known) telescope degradations. The correction factor found from performing the muon calibration on MC, $\varepsilon^{\mathrm{nom}}_\mu/\varepsilon^{\mathrm{deg}}_{\mu}$, should be linear with the ratio of the known input telescope optical degradations. Here, $\varepsilon^{\mathrm{nom}}_\mu$ is the muon efficiency obtained from using the calibration algorithm on MC of the telescopes at their nominal efficiency (defined as 100\%), whilst $\varepsilon^{\mathrm{deg}}_{\mu}$ is the muon efficiency obtained from using the calibration algorithm on MC of the telescopes at a degraded efficiency. The improvement in linear behaviour with the updated algorithm is clearly demonstrated in figure \ref{fig:mclinearity}.

\begin{figure*}
	\centering
	\includegraphics[width = \textwidth]{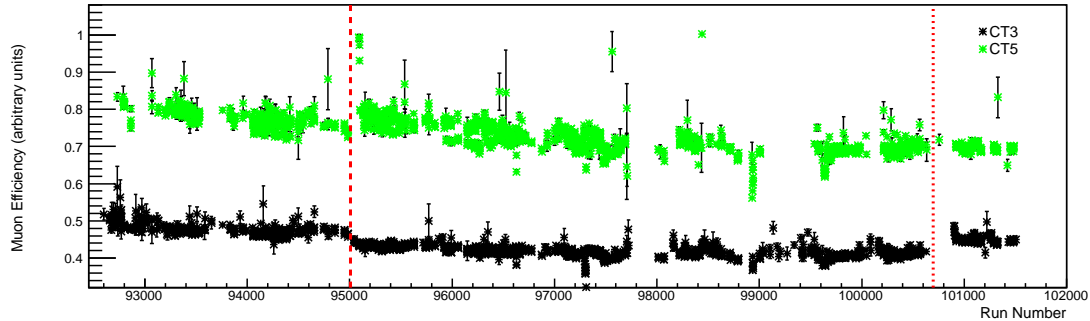}
	\caption{Evolution of the muon efficiency (in arbitrary units) of CT5 and one of the H.E.S.S. I telescopes during 2014. Discontinuous changes due to camera hardware adjustment in both telescopes, and cleaning of CT3 camera components are marked by dashed and dotted lines respectively.}
	\label{fig:MuonEffEvolution}
\end{figure*}

\noindent This generic algorithm is currently used on all five H.E.S.S. telescopes as part of the standard calibration chain. The evolution of the muon efficiency of CT5 and one of the H.E.S.S. I telescopes during 2014 is shown in arbitrary units in figure \ref{fig:MuonEffEvolution}. \noindent Although the overall efficiency evolution is mostly stable with a gradual degradation, the runwise efficiencies are still subject to some statistical variation. The merging of muon efficiencies over longer time periods, such as nightly, may reduce statistical variation to provide more stable estimates.
Discontinuous changes are discernible in both telescopes at run number 95000 due to camera hardware adjustment, and in CT3 at run number 100700, due to cleaning of some camera components. 

\section{Conclusions}
\label{sec:conclusion}
\noindent A generic muon algorithm has been implemented for H.E.S.S. II, with the flat distance across a telescope mirror determined by interpolation between a set of points defining the mirror edges. Smooth (circular or elliptical) mirror shapes are definable with a large number of points (unrestricted in principle). A 2D pixel log-likelihood fitting procedure replaces the previous 1D $\chi^2$ fit to the azimuthal muon intensity profile. These alterations have led to demonstrable improvements in the linearity of the correction factor with optical degradation. 

\noindent The development of flexible algorithms for telescopes of varying specifications is of increasing importance with the advent of the Cherenkov Telescope Array (CTA) \cite{bib:Acharya13}. We foresee no complications arising from the adaptation of this algorithm for arbitrary mirror shapes, whilst minor adjustments may be required for telescopes with Schwarzschild-Couder (dual mirror) optical design.

\section{Acknowledgements}
\noindent The support of the Namibian authorities and of the University of Namibia in facilitating the construction and operation of H.E.S.S. is gratefully acknowledged, as is the support by the German Ministry for Education and Research (BMBF), the Max Planck Society, the German Research Foundation (DFG), the French Ministry for Research, the CNRS-IN2P3 and the Astroparticle Interdisciplinary Programme of the CNRS, the U.K. Science and Technology Facilities Council (STFC), the IPNP of the Charles University, the Czech Science Foundation, the Polish Ministry of Science and Higher Education, the South African Department of Science and Technology and National Research Foundation, and by the University of Namibia. We appreciate the excellent work of the technical support staff in Berlin, Durham, Hamburg, Heidelberg, Palaiseau, Paris, Saclay, and in Namibia in the construction and operation of the equipment.



\end{document}